\documentclass[epsf,graphics,a4,12pt]{article}
\usepackage{epsfig}
\begin{document}
\baselineskip=18pt
\newcommand{\be}{\begin{equation}}
\newcommand{\en}{\end{equation}}
\begin{titlepage}
\begin{center}
{\hbox to \hsize{\hfill .}}
\bigskip
\vspace{6\baselineskip}
{\Large \bf  The Elliptic Solutions to the Friedmann equation and the Verlinde's Maps.}
\end{center}
\vspace{1ex}
\centerline{\large
Elcio Abdalla\footnote{eabdalla@fma.if.usp.br} and
L.Alejandro Correa-Borbonet\footnote{correa@fma.if.usp.br}}
\begin{center}
{Instituto de F\'{\i}sica, Universidade de S\~{a}o Paulo,\\
C.P.66.318, CEP 05315-970, S\~{a}o Paulo, Brazil}
\vspace{4ex}
\large{

{\bf Abstract}\\
}
\end{center}
\noindent
We considered the solutions of the Friedmann equation in several setups,
arguing that the Weierstra$\ss$ form of the solutions leads to connections
with some Conformal Field Theory on a torus. Thus a link with
the Cardy entropy formula is obtained in a quite natural way. The argument
is shown to be valid in a four dimensional radiation dominated universe with
a cosmological constant as well as in four further different Universes.
\\
\\
\vspace{9ex}
\hspace*{0mm} PACS number(s): 98.80.-k; 98.80.Jk; 02.40.-k; 02.90.+p
\vfill
\end{titlepage}

\newpage
\section{Introduction.}
Some time ago Verlinde\cite{Verlinde} considered the holographic principle
\cite{susskind} in a Fried\-mann-Robertson-Walker(FRW) universe in radiation
dominated phase. Such a radiation is considered to be described by a
Conformal Field Theory with a large central charge.
In that work a very interesting relationship between the FRW equations
controlling the cosmological expansion on the one hand and the formulas
that relate the energy and the entropy of the CFT on the other hand
was found. In particular three amazing
relations mapping the $D$-dimensional Friedmann equation
\be
H^{2}=\frac{16\pi G}{(D-1)(D-2)}\frac{E}{V}-\frac{1}{R^{2}}
\en
into the Cardy formula\cite{Cardy}
\be
S=2\pi \sqrt{\frac{c}{6}(L_{0}-\frac{c}{24})},
\en
very well known in two dimensional  CFT, has been set up, that is
\begin{eqnarray}
2\pi L_{0} & \Rightarrow & \frac{2\pi}{D-1}ER  \nonumber \\
2\pi \frac{c}{12} & \Rightarrow & (D-2)\frac{V}{4GR} \\
S  & \Rightarrow & (D-2)\frac{HV}{4G} \quad .\nonumber
\end{eqnarray}

 The scenario considered in \cite{Verlinde}  was that of a closed
radiation dominated FRW universe with a vanishing cosmological
constant. Such relations hold precisely when the holographic entropy bound
is saturated.
Within this context Verlinde proposed that the Cardy formula for $2D$ CFT
can be generalized to arbitrary spacetime dimensions. Such a
generalized entropy formula is known as the Cardy-Verlinde formula.

Later soon, this result was generalized and understood in several set ups
\cite{ivo,abda,wang,veneziano}. The CFT dominated universe has been described
as a co-dimension one brane, in the background of various kinds
of (A)dS black holes. In such cases, when the brane crosses the black hole
horizon, the entropy and temperature are expressed in terms of the Hubble
constant and its time derivative.
Verlinde's proposal has inspired a considerable activity shedding further
light on the various aspects of the Cardy-Verlinde formula. Nevertheless
there is still no answer to the
question about whether the merging of the CFT and FRW equations is a  mere
formal coincidence or, quoting Verlinde, whether this fact
``{\it strongly indicated that both sets of these
equations arise from a single underlying fundamental theory}''.
Moreover it is very striking that the Cardy formula obtained in the very
particular framework of two dimensional Conformal Field Theory gets
generalized as directly as proposed. One thus wonders whether there is no
essentially two-dimensional aspect overlooked in the whole calculation.

Aiming at a more complete picture of such a difficult task
we seek, in this paper, new connections that can be established between the
Friedmann equation and the Cardy formula. Our point of departure is the work
of Kraniotis and Whitehouse \cite{kraniotis}, where those authors found
solutions for the Friedmann equation that directly and naturally
connects the Friedmann-Robertson-Walker cosmology with the theory of modular forms and
elliptic curves \cite{koblitz}.

\section{Friedmann equation and Weierstra$\ss$  form .} \nonumber
In this section we apply the procedure shown in \cite{kraniotis} to the
radiation\footnote{In the reference \cite{kraniotis} the matter
dominated universe is studied.} dominated four-dimensional universe with $k=1$ and
non vanishing cosmological constant. The Friedmann equation takes the form
\be
H^{2}=\frac{8\pi
G}{3}\frac{C_{1}}{R^{4}}+\frac{\Lambda}{3}-\frac{1}{R^{2}} \label{eq:Frie}
\en
where the density is $\rho=E/V=C_{1}R^{-4}$.

Equation  (\ref{eq:Frie}) can be recast into the form
\be
(\dot {R})^{2}=-1+\frac{\Lambda}{3}R^{2}+\frac{8\pi G C_{1}}{3}R^{-2}.
\en
For the time variable we define the elliptic integral
\begin{eqnarray}
t & = & \int \left [-1+\frac{\Lambda}{3}R^{2}+\frac{8\pi G C_{1}}{3}R^{-2}
\right]^{-1/2} \, dR \\
  & = & \int R \left [-R^{2}+\frac{8\pi G C_{1}}{3}+
\frac{\Lambda}{3}R^{4}  \right]^{-1/2} \, dR \quad .
\end{eqnarray}
Now let us consider the auxiliar integral
\be
 u =\int  \left [-R^{2}+C_{2}+\frac{\Lambda}{3}R^{4} \right]^{-1/2}
\, dR   \quad ,\label{eq:ellip}
\en
with $C_{2}=\frac{8\pi G C_{1}}{3}$. We have
\be
t=\int R \, du  \quad .
\en

The integral (\ref{eq:ellip}) can be reduced to a cubic form upon introducing the
variable $X=\frac{1}{R^{2}}$, that is
\be
u=\int  \left [-X^{2}+C_{2}X^{3}+\frac{\Lambda}{3}X \right]^{-1/2}
\, (\frac{1}{2})dX \quad .
\en

We introduce the variable
$X=\frac{\xi+\frac{1}{12}}{\frac{C_{2}}{4}}$, leading us to the relation
\be
v=2u=\int \left [4\xi^{3}-(\frac{1}{12}-\frac{\Lambda C_{2}}{12})\xi-
\frac{1}{216}+\frac{\Lambda C_{2}}{12^{2}} \right]^{-1/2} \,d\xi \quad .
\en

This integral can be identified with the Weierstra$\ss$ normal form,
arising from the well known differential equation of the
Weierstra$\ss$ function $\wp$ \cite{abramowi,tate}, that is
\be
(\wp^{\prime})^2=4\wp^3-g_2 \wp-g_3 \quad .
\en
The Weierstra$\ss$ function $\wp(u)=\wp(u|\omega_{1},\omega_{2})=
\wp(u,\tau=\omega_{1}/\omega_{2})$
is an even meromorphic elliptic function of semiperiods
$\omega_{1}$, $\omega_{2}$.
Such semiperiods are given by the expressions
\be
\omega_{1}=\int^{\infty}_{e_{1}} \frac{dz}{\sqrt{4z^{3}-g_{2}z-g_{3}}}
\,\,\,\, , \,\,\,\,\, \omega_{2}=\int^{\infty}_{e_{3}} \frac{dz}{\sqrt{4z^{3}-g_{2}z-g_{3}}} \quad ,
\en
where $e_{1},e_{2},e_{3}$ are the roots of the cubic polynomial
$4z^{3}-g_{2}z-g_{3}$ with discriminat
\be
\Delta=g^{3}_{2}-27g^{2}_{3} \quad .
\en

Thus $\xi=\wp(v+\epsilon)$, where $\epsilon$ is a constant of integration.
Moreover, the cubic invariants are
\be
g_{2}=\frac{1}{12}-\frac{\Lambda C_{2}}{12}  \,\,\,\,\,\,\,\,\, {\rm
and}
\,\,\,\,\,\,\,\,\,
g_{3}=\frac{1}{216}-\frac{\Lambda C_{2}}{12^{2}} \quad .\label{eq:inva}
\en
Substituting back into $R$, we find
\be
R=\sqrt{\frac{8\pi C_{1} G}{12\wp(v+\epsilon)+1}}
\en
and
\be
t=\frac{1}{2}\int R \, dv =\frac{1}{2}\int \sqrt{\frac{8\pi C_{1} G}
{12\wp(v+\epsilon)+1}} \, dv \quad .
\en

These interesting results open us another door to the understanding of
the Verlinde's maps. The fact that we can obtain the Weierstra$\ss$
$\wp$-function from the Friedmann equation shows a new way of
interpreting the relation between
the Cardy formula and the Friedmann equation. This is due to the particular
properties of this function. First, it is known from the theory of
the elliptic curves \cite{eichler} that the topological equivalence of
an elliptic curve
with a torus is given by an explicit mapping involving the
Weierstra$\ss$ $\wp$-function and
its first derivative. This map is, in effect, a parametrization of the
elliptic curve by points in a ``fundamental parallelogram'' in the complex
plane. Another important property of this function is its non trivial
modular transformation properties \cite{eichler}. It is a meromorphic
modular form
of weight $2$, that is,
\begin{eqnarray}
\wp \left( \frac{u}{c\tau +d},\frac{a\tau +b}{c\tau +d}\right) &=&(c\tau
+d)^{2}\wp (u,\tau ),
\quad
\left(
\begin{array}{cc}
a & b \\
c & d%
\end{array}%
\right) \in {\bf SL}(2,Z).
\end{eqnarray}

From (\ref{eq:inva}) it is not hard to see that for $\Lambda=0$ we get
$\Delta=0$. This case is not studied because the solutions are not given
by elliptic functions and do not have modular properties.

We thus stress that the most important result arising out of
this approach is the correspondence established between  the given
Friedmann equation
and a particular torus through the parameter $\tau$. It is clear that $\tau$
is a function of the cosmological constant $\Lambda$ depending on the
dimension as well (this will become evident in the next section).

On the other hand, we know that when we have a CFT theory,
a partition function on a torus with modular parameter
$\tau$ can be introduced, being given by the expression
\be
Z(\tau)=Tr\, q^{L_{0}-c/24} \quad ,
\en
where $q\equiv e^{2\pi i\tau}$ ( for simplicity we have suppressed the
$\bar{\tau}$ dependence).

 Using the modular properties of this  partition function it is possible
to find \cite{Cardy,carlip} the density of states and consequently the Cardy
entropy formula
\be
S=2\pi \sqrt{\frac{c}{6}(L_{0}-\frac{c}{24})} \quad .
\en

 These facts lead to the chain of
connections: Friedmann equation $\rightarrow$ Weierstra$\ss$ equation
$\rightarrow$ Torus $\rightarrow$ CFT Partition Function
Z($\tau(\Lambda$, $D$))$\rightarrow$ Cardy Formula. Therefore, in this
context, Verlinde's  maps
are less intri\-guing. We actually do not obtain a direct definition of the
quantities appearing in Cardy-formula as a function of the quantities
appearing in the Friedmann equation, but the very existence of a Cardy formula
is natural, given the partition function on a torus.
\section{Further examples.}
In this section we extend the approach of the previous section to two
further cases.
First we study the case $D=6$, $k=1$, $\Lambda \neq 0$. Here the Friedmann
equation takes the form
\be
H^{2}=\frac{16\pi
G}{(6-1)(6-2)}\frac{C_{1}}{R^{6}}+\frac{2\Lambda}{(6-1)(6-2)}-\frac{1}{R^{2}}
\label{eq:seis} \quad .
\en

The time integral and the auxiliar integral are now defined, respectively, as:
\begin{eqnarray}
t & = & \int \left [-1+\frac{\Lambda}{10}R^{2}+\frac{4\pi G C_{1}}{5}R^{-4}
\right]^{-1/2} \, dR \\
  & = & \int R^{2} \left [-R^{4}+\frac{4\pi G C_{1}}{5}+
\frac{\Lambda}{10}R^{6}  \right]^{-1/2} \, dR
\end{eqnarray}
and
\be
u=\int  \left [-R^{4}+C_{4}+
\frac{\Lambda}{10}R^{6}  \right]^{-1/2} \, dR \quad ,
\en
where $C_{4}=\frac{4\pi G C_{1}}{5}$.

Introducing $X=\frac{1}{R^{2}}$ we get
\be
u=\int  \left [-X+C_{4} X^{3}+\frac{\Lambda}{10}
\right]^{-1/2} \,\frac{1}{2} dX \quad .
\en

After some manipulations we arrive at the Weierstra$\ss$ normal form
\be
v=C^{1/2}_{4}u=\int  \left [4 X^{3}-\frac{4X}{C_{4}}+\frac{2\Lambda}{5C_{4}}
\right]^{-1/2} \, dX \quad ,
\en
with cubic invariants
\be
g_{2}=\frac{4}{C_{4}}  \;\;\;\;\;\;\;\;
{\rm and} \,\,\,\,\,\,\,\,\,
g_{3}=-\frac{2\Lambda}{5C_{4}} \quad .
\en
Substituting back into $R$, we find
\be
R=\sqrt{\frac{1}{\wp(v+\epsilon)}} \quad ,
\en
and
\be
t=\frac{1}{C^{1/2}_{4}}\int R^{2} \, dv =\frac{1}{C^{1/2}_{4}}\int
\frac{dv}{\wp(v+\epsilon)} \,  \quad .
\en

In this case the solution keeps the modularity when $\Lambda=0$
because $\Delta \neq 0$. From (\ref{eq:seis}) it is clear that the
parameters $g_{2},g_{3}$ and $\tau$ depends on the dimension. But here
as in the four dimensional case we obtain a map from the six dimensional
Friedmann equation to a particular torus with modular parameter $\tau=\tau
(\Lambda, D )$. This forces us to reexamine the argument appearing in the
literature that the Cardy-Verlinde formula is actually
D-dimensional. In fact we get a torus with a different $\tau$.

Similar results apply also to the eight-dimensional universe with radiation
dominated and $\Lambda=0$, where we find a solution in terms of the
Weierstra$\ss$ function, with
\be
g^{(D=8)}_{2}=0  \,\,\,\,\,\,\,\,\, {\rm
and}
\,\,\,\,\,\,\,\,\,
g^{(D=8)}_{3}=\frac{4}{C_{6}} \quad ,
\en
where $C_{6}=\frac{8\pi G C_{1}}{21}$.

\section{Conclusion.}
The solution of the Friedmann equation in terms of the
Weierstra$\ss$ function $\wp(u)$ leads to a modular invariant
partition function defined on a torus obtained from that cosmological equation.
Thus the two dimensional character of the Cardy-Verlinde formula
becomes evident, since there we can define a CFT partition function,
computing the entropy following Cardy's prescription. This work thus proposes a
natural two dimensional space for Cardy-Verlinde formula, establishing a
framework for Verlinde's maps. Although the form of the map has
different disguises in different dimensions and frameworks, we have obtained
the results in at least four different and inequivalent situations. We
think that this results lead us further in the comprehension of the
physical implications of Verlindes's proposal, implying new features
about the mathematical structure of the cosmological equations.
\\
\\
{\bf ACKNOWLEDGMENT:}
This work was partially supported
by Funda\c{c}\~ao de Amparo \`a Pesquisa do Estado de
S\~ao Paulo (FAPESP), Conselho Nacional de Desenvolvimento
Cient\'{\i}fico e Tecnol\'{o}gico (CNPq).

\newpage

\end{document}